
\input phyzzx
\def\CR{C_{2R'}}

\def\bR{\bar R}

\def\nt{\tilde n}
\def\nR{N_{R'\bR'R}}

\def\de{\delta}

\def\th{\theta}

\def\YM{{${\rm YM}_2$}}

\def\d{\partial}

\Pubnum={CERN-TH.7243/94\cr
USC-94/008\cr
hep-th/9404192}
\date={April 1994}
\titlepage
\title{Density Correlation Functions in Calogero-Sutherland Models}
\bigskip
\author {
Joseph~A.~Minahan\footnote\star
{minahan@physics.usc.edu}}
\address{Department of Physics,  \break
University of Southern California, Los Angeles, CA 90089-0484 USA}
\andauthor{
Alexios P. Polychronakos\footnote\dagger
{poly@dxcern.cern.ch}}
\address{Theory Division, CERN\break
CH-1211, Geneva 23, Switzerland}
\bigskip
\abstract{
Using arguments from two dimensional Yang-Mills theory
and the collective coordinate formulation of the Calogero-Sutherland model,
we conjecture the dynamical density correlation function for coupling $l$
and $1/l$, where $l$ is an integer.  We present overwhelming evidence
that the conjecture is indeed correct.
}
\submit{Physical Review B, {\rm PACS numbers: 03.65.Ca, 05.30.Fk,
03.65.Fd}}
\vfill
\endpage

\def\NP{{\it Nucl. Phys.\ }}
\def\PL{{\it Phys. Lett.\ }}
\def\PRD{{\it Phys. Rev. D\ }}
\def\PRB{{\it Phys. Rev. B\ }}
\def\PRA{{\it Phys. Rev. A\ }}
\def\PRL{{\it Phys. Rev. Lett.\ }}

\def\JMP{{\it J. Math. Phys.\ }}

\def\JP{{\it J. Phys.\ }}

\def\MPL{{\it Mod. Phys. Lett. A\ }}

\def\ZETF{{\it Zh. Eksp. Teor. Fiz.}}

\REF\Calog{F.~Calogero, {\it Jour. of Math. Phys.}
{\bf10}, 2191 and 2197 (1969); {\bf 12}, 419 (1971).}
\REF\Suth{B.~Sutherland, \PRA {\bf 4}, 2019 (1971); \PRA {\bf5}, 1372 (1972);
\PRL {\bf34}, 1083 (1975).}
\REF\HalShas{F.~D.~M.~Haldane, \PRL {\bf 60}, 635 (1988);
B.~S.~Shastry,\PRL {\bf 60}, 639 (1988).}
\REF\SLA{ B.\ D.\ Simons, P.\ A.\ Lee, and B.\ L.\ Altshuler,
\PRL {\bf 70}, 4122 (1993);\PRL {\bf 72}, 64 (1994);
E.~Mucciolo, B.~Shastry, B.~Simons and B.~Altshuler, cond-mat/9309030.
}
\REF\FStat{A. Polychronakos, \NP {\bf B234} (1989) 597; F.D.M. Haldane,
\PRL {\bf 67} (1991) 937;
Y.S. Wu, University of Utah preprint, cond-mat/9402008, February 1994;
D.~Bernard and Y.~S.~Wu, cond-mat/9404025, April 1994.}
\REF\Hal{F.~D.~M.~Haldane, cond-mat/9401001, January 1994}
\REF\Shas{B.~S.~Shastry, cond-mat/9402065.}
\REF\Migdal{A.~Migdal, \ZETF {\bf 69} (1975) 810;
B.~Rusakov, \MPL {\bf5} (1990) 693.}
\REF\MPD{
Y.~Nambu, in ``From SU(3) to Gravity,'' pp.45-52
Gotsman, E.~(Ed.), Tauber, G.~(Ed.);
and \PRD {\bf 26} (1992) 2875 and references therein;
M.~Stone, \PRB {\bf 42} (1990) 399;
A.~Jevicki, \NP {\bf B376} (1992) 75;
J.~A.~Minahan and A.~ P.~Polychronakos,
\PL {\bf B312} 155, 1993;
M.~Douglas, RU-93-13 hepth/9303159;
M. Caselle, A. D'Adda, L. Magnea, and S. Panzeri, hep-th/9304015, 1993.
}
\REF\QCDS{A. Gorskii, and N. Nekrasov, \NP {\bf B414} (1994) 213;
J.~A. Minahan and A.~ P. Polychronakos,
Sep. 1993. {\it to appear in Phys. Lett. B}, hep-th/9309044}
\REF\Stan{R.\ P.\ Stanley, {\it Adv.\ Math.} {\bf 77}, 76 (1989).}
\REF\SJ{A.~Jevicki and B.~Sakita, \PRD {\bf 22} (1980) 467.}
\REF\AJL{I.~Andric, A.~Jevicki and  H.~Levine, \NP {\bf B215} (1983) 307.}
\REF\Dyson{F.~Dyson, \JMP {\bf3} (1962) 157.}
\REF\AB{I.~Andric and V.~Bardek, \JP {\bf A21} (1988) 2847.}
\REF\Gaud{M.~Gaudin, Saclay preprint SPhT-92-158.}
\REF\SuthSh{B.\ Sutherland and B.\ S.\ Shastry, \PRL {\bf 71}, 5 (1993).}
\REF\Forr{ P.\ J.\ Forrester, \PL {\bf A179}, 127 (1993).}
\REF\Ha{Z.~N.~C.~Ha, {\it to appear}}
\REF\LPS{F.~Lesage, V.~Pasquier and D.~Serban, Saclay preprint, April 1994}

There has been much recent interest in the Calogero-Sutherland Model (CSM)
of interacting fermions in one dimension[\Calog,\Suth].
One reason for the interest is that these models are related to quantum spin
chains, with long range interactions between the spins[\HalShas].
The CSM models also have a close
relation to random matrix theory[\Suth,\SLA] and fractional
statistics[\FStat,\Hal,\Shas].

The Hamiltonian of the CSM on the circle is given by
$$H={1\over 2mL'^2}\left(-\sum_i{\d\over\d\th_i^2}
+\half\sum_{i\ne j}{l(l-1)\over\sin^2(\th_i-\th_j)/2}\right)\eqn\CSMHam$$
The spectrum of $H$ has been long known[\Suth], but until
recently the dynamical correlation functions remained unknown.
The basic problem is
that while in principle it is possible to
calculate the eigenfunctions for the CSM[\Suth],
in practice it is a computational nightmare.  A big break in this
direction came last year, when Simons, Lee and Altshuler calculated the
{\it dynamical} two point density correlation function for three
special values of the coupling.  They exploited the fact that for these
values of the coupling, the CSM was related to random matrix
theory of orthogonal, unitary or symplectic matrices.

In this letter we will generalize the results of Simons Lee and Altshuler
for CSM couplings $l$ and $1/l$, where $l$ is any integer.  Our analysis
is based on some straightforward observations about the form
of the measure in the correlation functions.  Using arguments from
two-dimensional Yang-Mills theory (\YM) and a duality symmetry found in the
CSM, we find that the functional form of the
measure must satisfy a host of properties.  We find a function that
satisfies these properties and conjecture that it is unique.

Our starting point will be $U(N)$ \YM\ on the torus.
It is by now well known that its partition function is given by the
sum[\Migdal]
$$Z=\sum_{R'} e^{-LTg^2\CR}\eqn\partfn$$
where $R'$ are the possible $U(N)$ representations, $L$ and $T$ are the lengths
of the two circles comprising the torus, $g$ is the \YM\ coupling
and $\CR$ is the quadratic Casimir for representation $R'$.  $R'$ is described
by a Young tableau with $n_i$ boxes in row $i$, which satisfy the constraints
that $n_i\ge n_{i+1}$.  Since we are considering
$U(N)$, the number of boxes in a row can be negative. The Casimir is then
given by
$$\CR=\half\sum_i n_i(n_i+N-2i+1)=\half\sum_i\left(n_i+{N-2i+1\over2}\right)^2
-{(N^2-1)N\over24}.\eqn\casimir$$
Letting $p_i=n_i+(N-2i+1)/2$, then $p_i$ satisfies $p_i\ge p_{i+1}+1$.  From
this it is clear that
the partition function in \partfn\ is the same as that for $N$
nonrelativistic free fermions on a circle where the length of this circle,
$L'$,
and the fermion mass $m$ satisfy the relation $mL'^2=1/(g^2L)$[\MPD].  $T$ is
the Euclidean time in this problem.
Next insert a Wilson loop in representation $R$
along the Euclidean time direction of the torus.  This corresponds to
putting a fixed
color source at some point on the original \YM\ circle.  The
loop modifies the partition function to
$$Z_R=\sum_{R'}\nR e^{-LTg^2\CR}\eqn\partfnR$$
where $N_{R_1R_2R_3}$ is the number of singlets in the tensor product
$R_1\times R_2\times R_3$.
 If $R$ is the $(l-1)N$ symmetric representation  of $SU(N)$ then $\nR=1$ if
$p_i\ge l+p_{i+1}$.  Hence, the allowed states
are precisely the states found in the CSM with coupling $l$[\QCDS].

The fact that the CSM is related to \YM\ suggests that other
techniques from \YM\ might be applicable, in particular the string
interpretation. This is equivalent to a formulation in terms of
creation and annihilation operators.
To achieve this, perform the standard
redefinition of the wavefunction, $\Psi=\Delta^l f$, where $\Delta$ is the
Vandermonde for $z_i=e^{i\th_i}$.
The function $f$ is a combination of \YM\ characters,  satisfying[\Suth]
$$
H f = {1\over mL'^2}
\left({l \over 2} \sum_{i\neq j} {z_i + z_j \over z_i - z_j}
(z_i \d_i -
z_j \d_j ) f + \sum_i (z_i^2 \d_i^2 + z_i \d_i ) f\right) = E f
\eqn\H$$
where $\d_i = \d_{z_i}$.
We can classify the states
in terms of representations $R'$ of $U(N)$ (just as in \YM)
as $\chi(R',l)$ with energy $E(R',l)$. These can be thought as $l$-deformed
characters and Casimirs.  These are also referred to as Jack polynomials
in the mathematical literature[\Stan]. Note that $\chi(R,1)=\chi_R$ is
the usual $U(N)$ character.
$f$ can be expanded in terms of the Schur functions
$W_n = \sum_i z_i^n$, giving
$$
f = \sum_k f\{n_1,\dots n_k\} \prod_{i=1}^k W_{n_i} \equiv
\sum_k f\{n_1,\dots n_k\} |n_1,\dots n_k \rangle
\eqn\S $$
where $f\{n_i \}$ are expansion coefficients. Since here we will be interested
only in chiral states, all $n_i$ can be chosen positive. This is the
``string'' picture, where the state $|\{n_i\}\rangle$
is interpreted as a collection
of strings with winding numbers $n_i$.
Defining string creation and annihilation operators
$|n\rangle \equiv a_n^+ |0\rangle$ and $n |0> \equiv a_n |n\rangle$
which obey the commutation relations
$[ a_n , a_m^+ ] = n \delta_{m,n}$
and interpreting the action of \H\ on \S\ in terms of $a_n$ and $a_n^+$,
we get[\SJ,\AJL]
$$
H={1\over mL'^2}
\Bigl[\sum_{n>0} \left\{ (1-l) n a_n^+ a_n + l N a_n^+ a_n \right\}
+\sum_{m,n>0} \left\{ l a_m^+ a_n^+ a_{m+n} + a_{m+n}^+ a_m a_n
\right\} \Bigr]
\eqn\HH$$
The terms inside the double sum in \HH\ are respectively
a splitting term and a joining term.
This string theory
is nonlocal on the world sheet, due to the existence
of the term $n a_n^+ a_n$, proportional to the {\it square} of the length
of each winding string.

A duality relation[\Stan,\Gaud] between the models with parameters
$l$ and $1/l$ can be established using the Hamiltonian in \HH.
To this end, define the new ``dual'' operators
$\tilde a_n^+ = -{1 \over l} a_n^+$ and $\bar a_n = -l a_n$
satisfying {\it the same} commutation relations as the original ones.
In terms of the new operators, the Hamiltonian becomes
$$\eqalign{
H={-l\over mL'^2}
\Bigl[\sum_{n>0} \bigl\{ (1-{1 \over l}) n \tilde a_n^+& \tilde a_n
+ {1 \over l} (-lN) \tilde a_n^+ \tilde a_n \bigr\} \cr
&+\sum_{m,n>0} \bigl\{ {1 \over l} \tilde a_m^+ \tilde a_n^+ \tilde a_{m+n}
+ \tilde a_{m+n}^+ \tilde a_m \tilde a_n \bigr\}\Bigr]}
\eqn\Hdual$$
So, apart from an overall constant $-l$, the Hamiltonian in terms of the
new operators is identical to the old one but with $l$ turned into $1/l$
and $N$ turned into $-lN$. Therefore, any  solution
of the $1/l$ problem, expressed as a polynomial of creation operators
acting on the vacuum, will also be a solution of the $l$ problem
upon substitution of the dual operators in its expression.  But
$\tilde a_n^+$ creates the state $-l W_n$. Further, turning $W_n$ into $-W_n$
in the expression for the character of a representation $R'$ changes it into
the character of the dual representation $\tilde R'$,
where the tableau of $\tilde R'$ is that of $R'$ with rows and
columns interchanged. Hence, one has the duality relations
$$
\chi(R',l;\{W_n\}) = \chi(\tilde R' , {1 \over l}, \{-lW_n\}),
\qquad
E(R',l,N) = -l E(\tilde R' , {1 \over l}, -lN)
\eqn\duality$$
Changing the positive integer $N$ into $-lN$ may look unphysical;
however, $N$ appears simply as a parameter in the Hamiltonian (its
other manifestation is in nonperturbative effects). Therefore we can
formally take it to be $-lN$ in all expressions explicitly involving it.
For generic $l$, duality maps particles into holes in the pseudomomentum
spectrum.  Since $n$ particles can annihilate with $ln$ holes,
this justifies the mapping $l \to 1/l$.
Obviously,
duality is broken by nonperturbative effects, since particle excitations
can have arbitrarily high momentum while holes cannot be excited any
further than the bottom of the Fermi sea.

In order to calculate correlation functions we must know the inner product
of the wavefunctions in terms of the correct measure.
It can be shown that in the large-$N$ limit
$<n|m> = {n \over l} \delta_{m,n}$.
In fact, this relation implies that in this limit,
$a_n^\dagger = l a_n^+$ (since $<n|m> =$ $<0|(a_n^+)^\dagger a_m^+ |0>$).
The Hamiltonian \HH\ is Hermitian under this
measure, and therefore the eigenstates are orthogonal with respect
to this measure.
To obtain the exact measure with direct calculation is nontrivial for
general $l$.
The alternative is to determine it
indirectly, through the jacobian of the change of variables from the
old wavefunction $\Psi (\theta_i )$ to the new one, viewed as a function
of the Schur variables $W_n$.
The evaluation of this jacobian is achieved
with standard collective field theory techniques Fourier transformed to
momentum space[\Dyson,\AB,\AJL].
We simply quote the result (appropriately modified for the periodic
Sutherland problem)
$$
\ln J = \sum_{n>0} l{W_{-n} W_n \over n} + (l-1) \sum_{k=2}^\infty
(-1)^k {1\over k(k-1)N^{k-1}} \sum_{\sum_{n_i}=0} W_{n_1} \cdots W_{n_k}
\eqn\meas$$
This result is exact perturbatively to all
orders in $1/N$,
and it contains an infinite number of vertices, of increasing order in $1/N$.
The inner product of any two states expressed as functionals of
$W_n$, is given by integrating over all $W_n$ with the measure
$J/\int J \prod_n dW_n$. In particular, the overlap $<n_1 \dots n_k | n_{k+1}
\dots n_{k+m}>$ is a $k+m$-point correlation function of an
interacting one-dimensional field theory with action \meas.
Since \meas\ is invariant under $W_n \to -l W_n$, $N \to -lN$, $l \to 1/l$,
the norm and the inner product of the states is invariant under duality
transformations.
The leading quadratic part of \meas, of order $N^0$, reproduces
the leading-order result
for the overlap $\langle n|m\rangle$. Since, however, we are
interested in the thermodynamic limit where the particle density remains
finite
we would like to compute the result to all orders in $1/N$.

\YM\ provides a useful framework for determing
selection rules in the density correlation functions.
The density operator for a system of $N$
fermions is given by
$$\rho(u)=(L')^{-1}\sum_i \de(u/L'-\th_i)=
(L')^{-1}\sum_n\sum_i e^{-inu/L'}e^{in\th_i}.\eqn\densityI$$
But the sum over $i$ is just the trace of an $N$ by $N$ unitary matrix whose
eigenvalues are $e^{i\th_i}$.
Hence $\rho(u)$ is
$\rho(u)=(2\pi L')^{-1}\sum_n e^{-inu/L'}W_n$.
The density correlation function is found by inserting onto the torus
two spatial Wilson loops, $\rho(u)$ and $\rho(v)$, which are
separated by Euclidean time $T$
and which cross the Wilson loop, $R$.
The loops divide the surface of the torus into two regions and to find
the correlations on the ground-state the area of one of these regions
is taken to infinity.
Using arguments of Migdal and Rusakov[\Migdal],
the density correlation function reduces to
$$\eqalign{ \langle\rho(u,t)\rho(v,0)\rangle=
{1\over (2\pi L')^2}\sum_n &e^{-in(u-v)/L'}\int dUdVdW\sum_{R'}d_{R'}d_{G}
\chi_{G}(UWV^\dagger W^\dagger)\cr
\chi_{R'}(UV^\dagger)
&\times\chi_R(W)\tr(U^n)\tr(V^{n\dagger})
e^{-tLC_{R'}}}
\eqn
\dcorr$$
$G$ is the so called ``staircase'' with $l-1$ boxes per step and
$d_{R'}$ is the dimension of representation $R'$. The integrals are over
the $U(N)$ group volume.
The Schur function $W_n$ can be expressed as the sum of $U(N)$ characters
$$W_n=\chi_{[n]}(U)-\chi_{[n-1,1]}(U)+\chi_{[n-2,1,1]}(U) + ...
(-1)^n\chi_{[1,1,1,..1]}(U)
\eqn\schurtochar$$
where the numbers in the square brackets correspond to the number of boxes
in each row of the corresponding tableau.
The only representations that appear in $W_n$ are the ``corner''
tableaux, that is, those that only have boxes in the first row or first column.

In principle, the integration \dcorr\ can be carried out,
but in practice it is highly tedious.  The result of each integral will
be given by some Clebsch-Gordan coefficients.
However, the contribution from $R'$ is zero in \dcorr\
if the tensor product of $G$ with all possible corner representations
does not contain $R'$.
This then leads to our selection rules, which are a generalization of
those in [\Shas].
Define the representation $G_\rho$ as the representation obtained
by adding the boxes in row $i$ of $\rho$ to row $i$ of $G$.
Then it is straightforward to see that the allowed representations
$R'$ are given by $G_\rho$,
where $\rho$ can have more than one box in each of the first $l$ columns,
but has at most one box after that.

The correlation functions of greatest interest are those
found in the large radius limit, where the overall fermion density is
kept fixed.  This implies that $N/L'$ is fixed  and therefore
$g^2N^2$ is fixed.  This differs
from the more traditional large $N$ limit of Yang-Mills theory where $g^2N$
is fixed.  With this new limit, a
$U(N)$ representation which has
order $N$ boxes in its tableau has a finite
energy in the large $N$-limit.
It is useful to define new variables $x_i=n_i/N$ and $y_i=\nt_i/N$, where
$n_i$ is the number of boxes in the $i^{\rm th}$ row and $\nt_i$ is
the number in the $i^{th}$ column.  To leading order in $1/N$, the
energy of a particular representation $R'$ is 
$E(x_1,y_1..y_l)={g^2N^2L\over2}\Bigl(x_1(x_1+l)-l\sum_i^ly_i(1-y_i)\Bigr)$.
 From the selection rules,
the density correlation function is given by $l+1$ sums,
where the first $l$ are the sums over the number of boxes
in the first $l$ columns, and the other sum is over the number of boxes
in the first row.
In the scaling limit, the sums become integrals and the integration limits
are $0<x_1<\infty$ and $0<y_l<y_{l-1}<y_{l-2}..y_1<1$.
It is easy to understand the nature of these sums by considering the
Bethe ansatz picture for the CSM[\Suth,\SuthSh].  The ground state
has $l-1$ empty sites between each filled site.
A representation $R'$ corresponds to the situation
where fermion 1 moves over $n_1$ sites, fermion 2 moves over $l$ sites,
as do the next $\nt_l-2$ fermions.  Then the next fermion moves over only $l-1$
sites, and the logic is continued all the way down to fermion $\nt_1$ which
moves over 1 site.
This is then equivalent to exciting one particle out of the fermi sea and
leaving $l$ holes.  $ly_i-l/2$ is the momentum of
$i^{\rm th}$ hole, and $x_1+l/2$ is the momentum of the particle.
Hence the correlation function should be given by
$$\eqalign{\langle\rho(u,t)\rho(v,0)\rangle=&
{N^2\over(2\pi L')^2}{\rm Re}
\int_0^\infty dx_1\int_0^1dy_1\cdots\int_0^{y_{l-1}}dy_{l}
\mu(x_1,y_1..y_l;l)\cr
&e^{iN(x_1+\sum_iy_i)(u-v)/L'}e^{-tN^2[x_1(x_1+l)-l\sum_iy_i(1-y_i)]/(2mL'^2)}}
\eqn\corrfn$$

It is now just a question of determining
$\mu(x_1,y_1..y_l;l)$.
To this end, we make the following observations:
({\it i}) The measure should be invariant under an interchange of
the $y_i$, since this just corresponds to interchanging the momenta
of the holes.  ({\it ii}) The measure should be invariant under $x_1\to-l-x_1$,
$y_i\to1-y_i$, since this corresponds to a parity transformation.  ({\it iii})
The integrand should vanish as $y_i\to y_{j}$, since there can not be two
holes in the same place.  ({\it iv}) Since $\langle W_{-n} W_n\rangle=N$ if
$n\ge lN$, the measure is nonzero and finite as $x_1\to\infty$.

For a particular eigenfunction of the Hamiltonian, the contribution to
the measure is proportional to
$\langle W_n|\psi\rangle\langle\psi|\psi\rangle^{-1}\langle\psi|W_n\rangle$.
There are some known eigenfunctions for the CSM,  such as the $n$-antisymmetric
wave-functions, which are given by
$$\psi_n=\sum_{n\ {\rm partitions}}\prod_{k=1}^n {(W_k)^k\over (n_k)!}.$$
The sum is over all partitions of $n$.
We can then make the additional observations: ({\it v}) Using the Jacobian
in \meas\ the contribution of $\psi_n$ to the
correlation function is $C(l)n^{(1-1/l)}[1+(l-1)n/(lN)]+{\rm O}(1/N^2)$, where
$C(l)$ is an $l$ dependent constant.  ({\it vi}) By duality, the contribution
of the `symmetric' representation, $\widetilde\psi_n$ is
$C(1/l)n^{(1-l)}[1+(l-1)n/(lN)]+{\rm O}(1/N^2)$.
These six constraints are highly restrictive and the only function that
we could find that satisfies them is
$$\mu(x_1,y_1..y_l;l)=
C^{-1}{\prod_{i<j}(y_i-y_j)^{2/l}\over{\prod_i(y_i(1-y_i))^{1-1/l}}}
{(x_1(x_1+l))^{l-1}(x_1+\sum_i y_i)^2\over\prod_i(x_1+ly_i)^2}
\eqn\result$$
That this function satisfies the first four constraints is obvious.  The
last two require further explanation.  The region of the integrand that
comes from the antisymmetric wave function is $1/N\sim y_i<<y_1<<1$
for $i>1$,
and $1/N\sim x_1<<y_1$.  Since we have replaced sums by integrals,
the $l+1$ integrals are integrated over a region of width $1/N$.
We then find that the contribution of $\psi_n$
is $\sim (N)^{1-1/l}y^{1/l}/(1-y)^{1-1/l}$, and the
behavior agrees with ({\it v}).
Likewise the contribution from $\widetilde\psi_n$
comes from $1/N\sim y_i<<x_1<<1$, hence after integrating the variables
over a width $1/N$, one finds the behavior is
$\sim  N^{1-l}x_1^{1-l}(l+x_1)^{l-1}$,
which agrees with ({\it vi}).

In fact there are a few more checks that we can make.  First,
\corrfn\ and \result\  agree with the previously known cases of $l=1$
and $l=2$[\SLA].
Second, note that the
contribution from $W_n$ comes from the region $\tau=x_1+\sum_i y_i$, where
$n=\tau N$.  We can pick off this contribution by inserting
$\de(\tau -x_1-\sum_i y_i)$ into the integrand.
If $\tau \ge l$, we can do the $x_1$ integration, leaving the integrals over
$y_i$ saturated.
But at equal time, we know that $\langle W_n W_{-n}\rangle=N$ for $n\ge lN$.
Therefore, if \result\ is correct, then it must be true that
$${1\over l!}\int_0^1\prod_i{dy_i\over(y_i(1-y_i))^{1-1/l}(\tau +ly_i-Y)^2}
\prod_{i<j}(y_i-y_j)^{2\over l}[(\tau -Y)(\tau -Y+l)])^{l-1}\tau
^2=C\eqn\tind$$
where $Y=\sum_iy_i$ and $C$ is a constant.
The constant is determined by
taking $\tau \to\infty$, in which case \tind\ reduces to the Selberg integral
and $C$ is given by
$$C={1\over 2l!}\left({l\over\Gamma(1/l)}\right)^l\prod_{k=1}^l\Gamma(k/l)
\eqn\Selberg$$
We have been unable to prove \tind\ directly, but we computed the $l=3$
case numerically and were able to show that it agreed with $C$ up to six
significant digits all the way down to $\tau =3.5$.  Finally, we observe
that as $x_1\to\infty$, the measure reduces to the measure recently found
by Forrester for the two-particle Green function[\Forr,\Hal].
This is sensible, since
sending the particle to infinite momentum should be equivalent to removing
it altogether.

Using the duality symmetry we can find the correlation function for
the case $1/l$.  In this case, we have $l$ integrals over variables $x_i$
that run from $0$ to $\infty$ and one integral over $y_1$ from $0$ to $1$.
The correlation function is then found by substituting $-N/l$ for $N$,
$-ly_1$ for $x_1$ and $-lx_i$ for $y_i$ in the
measure  and energy.
The result for $l=1/2$ agrees with the result found previously[\SLA].

{\it Note added}  After this work was completed, we learned that Ha has
been able to prove these results using properties of the
Jack polynomials[\Ha].  After this paper was submitted we received
another preprint proving these results[\LPS]

\ack{The research of J.A.M.~was supported in part by D.O.E.~grant
DE-FG03-84ER-40168. J.A.M. thanks CERN and M. Fowler at the University
of Virginia for their hospitality while this work was in progress.}

\refout
\end